\documentclass[aip,reprint]{revtex4-1}

\usepackage{graphics}

\begin{document}


\title{Absolute calibration of photon-number-resolving detectors with an analog output using twin beams}

\author{Jan {Pe\v{r}ina,~Jr.}}
\email{jan.perina.jr@upol.cz}
\affiliation{RCPTM, Joint Laboratory
of Optics of Palack\'{y} University and Institute of Physics AS
CR, 17. listopadu 12, 77146 Olomouc, Czech Republic.}

\author{Ond\v{r}ej Haderka}
\affiliation{Institute of Physics AS CR, Joint Laboratory of
Optics of Palack\'{y} University and Institute of Physics AS CR,
17. listopadu 12, 772 07 Olomouc, Czech Republic.}

\author{Alessia Allevi}
\altaffiliation{CNISM UdR Como, I-22100 Como, Italy.}
\affiliation{\hbox{Dipartimento di Scienza e Alta Tecnologia,
Universit\`a degli Studi dell'Insubria, I-22100 Como, Italy.}}

\author{Maria Bondani}
\altaffiliation{CNISM UdR Como, I-22100 Como, Italy.}
\affiliation{Istituto di Fotonica e Nanotecnologie, CNR-IFN,
I-22100 Como, Italy.}

\begin{abstract}
A method for absolute calibration of a photon-number resolving
detector producing analog signals as the output is developed using
a twin beam. The method gives both analog-to-digital conversion
parameters and quantum detection efficiency for the photon fields.
Characteristics of the used twin beam are also obtained. A
simplified variant of the method applicable to fields with high
signal to noise ratios and suitable for more intense twin beams is
suggested.
\end{abstract}


\maketitle

Detection of photons by detectors sensitive to electric-field
intensities represents the basic and by far the most important
diagnostic tool for optical fields \cite{Mandel1995}. That is why
quality and capabilities of optical detectors have continuously
attracted a great deal of attention in the whole history of modern
optics. Construction of detectors with single-photon resolution
(photomultipliers, semiconductor single-photon counting modules)
represented a milestone in this effort. The ability to build
photon-number resolving detectors (PNRD) reached some years ago
was undoubtedly the next milestone. Due to advanced technologies
available that time, several kinds of PNRDs have been designed.
Some of them, including optical-fiber-loop detectors
\cite{Haderka2004,Rehacek2003,Fitch2003} 
and intensified CCD cameras \cite{Haderka2005a}, 
produce a digitized signal that directly determines the number of
detected photons called photo-electrons. However, several kinds of
detectors give only analog outputs and so this output has to be
somehow mapped onto integer photo-electron numbers.
Super-conducting bolometers \cite{Miller2003,Fukuda2011},
transition-edge sensors \cite{Avella2011}, silicon
photomultipliers \cite{Jiang2007,Afek2009,Ramilli2010,Allevi2010}
and hybrid photo-detectors (HPD) \cite{Bondani2009,Allevi2010} can
be mentioned as typical examples. Calibration of such detectors
inevitably requires at least two constants, one giving the
probability of detecting a photon by creating a photo-electron
[absolute quantum detection efficiency (QDE)] and one providing
the mapping between the analog signal and photo-electron numbers.
Reliable determination of these constants then opens the door for
measuring quantum properties of optical fields at the level of
individual photons \cite{Perina1991}. As a result it gives
qualitatively better tools for the experimental analysis of
nonclassical fields.

There exist classical approaches for absolute detector calibration
for both kinds of detectors. However, an elegant calibration
method based on weak entangled two-photon fields
\cite{Malygin1981} and developed for single-photon resolving
detectors over more than thirty years \cite{Migdall1999} has
provided an inspiration also for PNRDs with digitized outputs.
Contrary to weak two-photon fields, the method naturally uses
more-intense twin beams (TB) with mean photon-pair numbers larger
than one \cite{Brida2010,PerinaJr2012a}. Here, we further
generalize the method to account for analog outputs of the
remaining kinds of PNRDs. Similarly to the approach presented in
\cite{PerinaJr2012a}, the method also provides the characteristics
of the used TB.
We note that another generalization of the original method has
been given mixing the analyzed field with a heralded single-photon
field \cite{Avella2011}.

We demonstrate the method by considering a suitable TB measured
simultaneously by two HPDs \cite{Bondani2009,Allevi2010}. HPDs
provide resolution for small numbers of detected photo-electrons.
Current coming from the semiconductor cathode of an HPD shows
several peaks that can be assigned to different photo-electron
numbers. After amplification and conversion of the measured real
values of voltages into (arbitrarily scaled) dimensionless
voltages, we obtain pairs ($ v_s, v_i $) of dimensionless voltages
characterizing simultaneous detections in the signal and idler
beams. By repeating the measurement $ N $ times, we obtain the
moments of the ensemble of real values of voltages ($ v_{s,j},
v_{i,j} $):
\begin{eqnarray}   
 \langle v_s^k v_i^l \rangle_a = \frac{1}{N}
 \sum_{j=1}^{N} v_{s,j}^k v_{i,j}^l , \hspace{5mm} k,l=0,1,\ldots
\end{eqnarray}
Symbol $\langle \rangle_a $ denotes a mean value taken over the values of
measured quantities.
As usual for reasonable numbers $ N $ of measurement repetitions,
only the considered first and second moments are reached with
sufficient precision \cite{PerinaJr2012a,PerinaJr2013a}.

Quantum nature of the detected light composed of photons, however,
requires the introduction of discrete quantities instead of the
real-valued voltages $ v_{s,j} $ and $  v_{i,j} $. That is why we
have to assign a certain interval of voltages for each state with
a fixed number of detected photo-electrons. To do this we define
voltage windows of widths $ \delta v_s $ and $ \delta v_i $ for
the signal and idler voltage axes, respectively, and assign
discrete values $ m_{c,j} $ to the measured values $ v_{c,j} $ ($
c=s,i $) by the formula
\begin{equation}   
 m_{c,j} = {\rm mod} \left[ v_{c,j}/\delta v_c + 1/2
 \right] . 
\label{2}
\end{equation}
Photo-electron histogram $ f_{\delta v}(m_s,m_i) $ depending on
windows' widths $ \delta v_s $ and $ \delta v_i $ can then be
built from the ensemble of pairs ($ m_{s,j},m_{i,j} $).

Using a classical approach, windows' widths $ \delta v_s $ and $
\delta v_i $ can be determined invoking a special calibration
procedure \cite{Andreoni2009,Bondani2009}. The measurement of TBs,
however, suggests an alternative way for their determination. As
TBs contain predominantly photon pairs, values of voltages $
v_{s,j} $ and $ v_{i,j} $ are correlated despite final detectors'
QDEs $ \eta_s $ and $ \eta_i $ in both detection arms. So also
signal and idler photo-electron numbers $ m_s $ and $ m_i $ are
correlated. This correlation is quantified by covariance $
c_{m,\delta v} $, which depends on widths $ \delta v_s $ and $
\delta v_i $:
\begin{equation}   
 c_{m,\delta v} = \langle \Delta m_s \Delta m_i\rangle /
 \sqrt{ \langle (\Delta m_s)^2 \rangle \langle (\Delta m_i)^2 \rangle}.
\label{3}
\end{equation}
where $ \Delta m = m - \langle m\rangle $ and $ \langle m_s^k
 m_i^l \rangle = \sum_{j=1}^{N} m_{s,j}^k m_{i,j}^l / N $, $
k,l=0,1,\ldots $. Covariance $ c_{m,\delta v} $ attains its
maximum for certain values of windows' widths $ \delta v_s $ and $
\delta v_i $. These values are optimal as they maintain the
pair-wise character of detected twin beams in the best possible
way.

\begin{figure}         
 \resizebox{0.9\hsize}{!}{\includegraphics{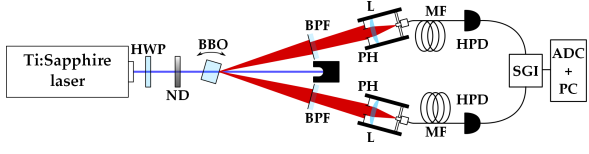}}
  \caption{Scheme of the experimental setup; HWP - half-wave plate,
  ND - neutral-density filter,
  BBO - BaB$ {}_2 $O$ {}_4 $ nonlinear crystal, BPF - band-pass filter, PH - pinhole, L - lens, MF -
  multi-mode fiber, HPD - hybrid photo-detector, SGI - synchronous-gated integrator, ADC -
  analog-to-digital converter.}
\label{fig1}
\end{figure}

\begin{figure}         
 \resizebox{0.99\hsize}{!}{\includegraphics{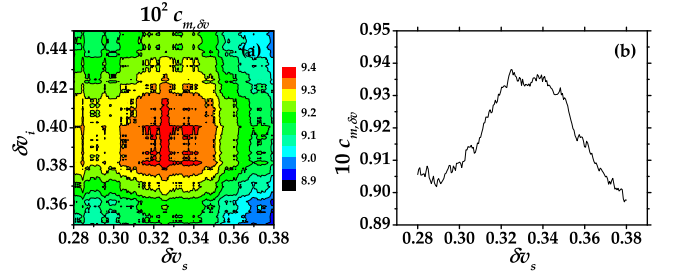}}
  \caption{(a) Topo graph of covariance $ c_{m,\delta v} $ as a function
  of windows' widths $ \delta v_s $ and $ \delta v_i $. (b)
  Covariance $ c_{m,\delta v} $ depending on $ \delta v_s $
  assuming $ \delta v_i = r \delta v_s $, $ r = 1.17 $. In both
  cases, the maximum of $ c_{m,\delta v} $ ($9.37 \times
  10^{-2} $) is reached for
  $ \delta v_s = 0.326 $ and $ \delta v_i = 0.375 $.}
\label{fig2}
\end{figure}

In the experiment shown in Fig.~\ref{fig1}, TB was generated in a
type I BaB$ {}_2 $O$ {}_4 $ crystal (cut angle $\vartheta_c=48
\deg$) pumped by the third harmonics (at 266 nm) of a
cavity-dumped Ti:Sapphire laser. 
The pulses were delivered at frequency 11~kHz to match the maximum
repetition rate of the detection apparatus. The polarization of
pump beam was adjusted by means of a half-wave plate. The
collection of non-collinear frequency-degenerated (at 532~nm)
signal and idler fields was performed 20~cm beyond the nonlinear
crystal using bandpass filters followed by multi-mode fibers
(600-$\mu$m-core diameter) and two HPDs (mod.~R10467U-40,
Hamamatsu, Japan) \cite{Andreoni2009,Bondani2009}. The output of
each HPD was amplified, synchronously integrated and digitized.
The experimental results confirmed a convex dependence of
covariance $ c_{m,\delta v} $ on widths $ \delta v_s $ and $
\delta v_i $. This dependence is demonstrated in
Fig.~\ref{fig2}(a) for typical experimental data giving $ \langle
v_s \rangle_a = 0.534 $, $ \langle v_i \rangle_a = 0.545 $, $
\langle (\Delta v_s)^2 \rangle_a = 0.557 $, $ \langle (\Delta
v_i)^2 \rangle_a = 0.572 $, and $ \langle \Delta v_s \Delta v_i
\rangle_a = 0.053 $.
If values of QDE's $ \eta_s $ and $ \eta_i $ are close to each
other and we measure a TB with small amount of noise photons, the
mean values of measured voltages depend linearly on detector
amplification and so also on windows' widths. We can thus assume
that
\begin{equation}   
 \delta v_i / \delta v_s \approx \langle v_i\rangle_a /
  \langle v_s\rangle_a \equiv r .
\label{4}
\end{equation}
This further simplifies the analysis and improves precision in the
determination of windows' widths $ \delta v_s $ and $ \delta v_i $
[see Fig.~\ref{fig2}(b)].

Once the windows' widths $ \delta v_s $ and $ \delta v_i $ are
fixed, the QDEs $ \eta_s $ and $ \eta_i $ together with the joint
signal-idler photon-number distribution $ p(n_s,n_i) $ [PND]
describing the generated TB in front of the detectors can be
determined using the experimental photo-electron histogram $
f(m_s,m_i) $. Following the approach described in
\cite{PerinaJr2012a}, TBs are considered as composed of three
independent components characterizing ideal paired field (index $
p $), signal noise photon field ($ s $) and idler noise photon
field ($ i $). The components are assumed in multi-mode thermal
states given for the photon-paired field by the fundamental theory
\cite{Perina2005,Perina2006}. %
As such they are composed of $ M_c $ independent equally-populated
modes with $ b_c $ mean photon (-pair) numbers per mode ($ c=p,s,i
$). We note that if the number $ M $ of modes is sufficiently
large (usually $> 5$), a field spectral profile composed of
differently populated modes plays only a negligible role in the
determination of PND \cite{Perina1991,Goldschmidt2013}. The
corresponding PND $ p(n_s,n_i) $ can be written as a two-fold
convolution comprising three Mandel-Rice distributions
\cite{Perina1991,PerinaJr2012a}:
\begin{eqnarray}  
 p(n_s,n_i) &=& \sum_{n=0}^{{\rm min}[n_s,n_i]} p(n_s-n;M_s,b_s)
   \nonumber \\
 & & \mbox{} \times p(n_i-n;M_i,b_i) p(n;M_p,b_p) \; ,
\label{5}
\end{eqnarray}
where $ p(n;M,b) = \Gamma(n+M) / [n!\, \Gamma(M)] b^n/(1+b)^{n+M}
$; $ \Gamma $ is the $ \Gamma $-function. Numbers $ M_c $ of modes
and mean photon (-pair) numbers per mode $ b_c $ can be derived
from the first and second photon (-pair) moments of these
components using the expressions for thermal fields.
\begin{eqnarray}   
 M_c = \frac{ \langle n_c \rangle^2 }{ \langle (\Delta n_c)^2\rangle -
  \langle n_c \rangle }, \hspace{4mm}
 b_c = \frac{\langle (\Delta n_c)^2 \rangle }{ \langle n_c
  \rangle } -1 .
\label{6}
\end{eqnarray}

Photon (-pair) moments written in Eq.~(\ref{6}) and experimental
photo-electron moments in Eq.~(\ref{3}) fulfil the relations
established by photo-detection theory \cite{Perina1991}. In this
theory, QDEs $ \eta_s $ and $ \eta_i $ are introduced as detector
parameters and the needed relations are derived as ($c=s,i$):
\begin{eqnarray}   
 \eta_c \left[ \langle n_p\rangle + \langle n_c\rangle \right] =
  \langle m_c\rangle, & &
 \eta_s\eta_i \langle (\Delta n_p)^2 \rangle =
  \langle \Delta m_s \Delta m_i \rangle ,   \nonumber \\
 \eta_c^2 \Biggl[ \langle (\Delta n_p)^2 \rangle + \langle (\Delta n_c)^2 \rangle
   &+& \frac{1-\eta_c}{\eta_c}\left( \langle n_p\rangle +
  \langle n_c\rangle \right) \Biggr] \nonumber \\
  &=& \langle (\Delta m_c)^2 \rangle \; .
 \label{7}
\end{eqnarray}
In five Eqs.~(\ref{7}), there occur eight unknown parameters: six
parameters of TBs and two QDEs. We express the photon moments in
the terms of photo-electron moments and mean photon-pair number $
\langle n \rangle_p $, which is taken as an independent parameter
($c=s,i$):
\begin{eqnarray}   
\langle n_c\rangle = \frac{\langle m_c\rangle}{\eta_c} - \langle
   n_p\rangle , \hspace{2mm}
 \langle (\Delta n_p)^2 \rangle = \frac{\langle \Delta m_s  \Delta m_i
  \rangle}{\eta_s\eta_i}, \nonumber \\
\langle (\Delta n_c)^2 \rangle = \frac{ \langle (\Delta m_c)^2
  \rangle}{ \eta_c^2} - \frac{ \langle \Delta m_s  \Delta m_i
  \rangle}{ \eta_s\eta_i} -\frac{1-\eta_c}{\eta_c^2} \langle m_c \rangle .
  \hspace{2mm}
\label{8}
\end{eqnarray}
Values of the remaining three unknown parameters, $ \langle n
\rangle_p $, $ \eta_s $ and $ \eta_i $, can be found by minimizing
the declinations between the theoretical and experimental
photo-electron histograms $ p_m $ and $ f $ quantified by function
$ D $:
\begin{equation}  
  D = \sqrt{ \sum_{m_s,m_i=0}^{\infty} \left[p_m(m_s,m_i) -
  f(m_s,m_i)\right]^2 } .
\label{9}
\end{equation}

The theoretical photo-electron histogram $ p_m $ can be obtained
from the PND $ p $ in Eq.~(\ref{6}) provided that HPDs are
described by the Bernoulli distribution $ T_k $
\cite{PerinaJr2012}:
\begin{eqnarray}   
 & & \hspace{-2mm} p_c(m_s,m_i) = \sum_{n_s,n_i=0}^{\infty} \hspace{-2mm} T_s(m_s,n_s)
   T_i(m_i,n_i) p(n_s,n_i),
\label{10}   \\
 & & \hspace{-2mm} T_k(m,n) = \left( \begin{array}{c} m \cr n \end{array}\right)  \eta_k^m
  \left( 1-\eta_k\right)^{m-n} , \hspace{5mm} k=s,i.
\label{11}
\end{eqnarray}

\begin{figure}         
 \resizebox{0.97\hsize}{!}{\includegraphics{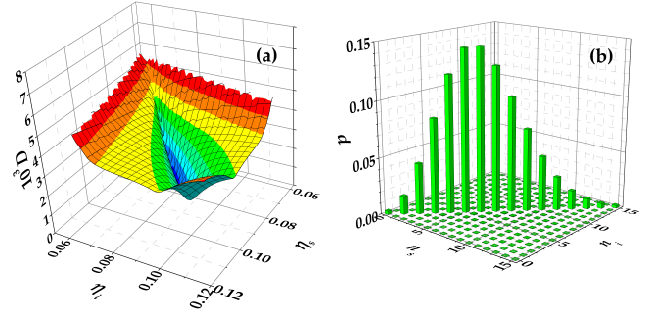}}
  \caption{(a) Minimum of declination function $ D $ determined
  over the allowed values of $ \langle n_p\rangle $ in dependence
  on QDEs $ \eta_s $ and $ \eta_i $. In areas close to the $ \eta_s $
  and $ \eta_i $ axes, Eqs. (\ref{7}) have no solution. In (b), signal-idler
  PND $ p(n_s,n_i) $ for optimal values of QDEs
  $ \eta_s = 0.085 $ and $ \eta_i = 0.086 $ is shown.}
\label{fig3}
\end{figure}

The declination function $ D $ in Eq.~(\ref{9}) depends on $
\langle n \rangle_p $, $ \eta_s $ and $ \eta_i $ and attains its
minimum that identifies suitable values for the unknown
parameters. This is documented in Fig.~\ref{fig3}(a) for the
experimental data analyzed in Fig.~\ref{fig2} assuming windows'
widths $ \delta v_s $ and $ \delta v_i $ maximizing the covariance
$ c_{m,\delta v} $. According to the graph in Fig.~\ref{fig3}(a),
the minimum of $ D $ is reached for $ \eta_s = 0.085\pm 0.002  $
and $ \eta_i = 0.086\pm 0.002 $. We note that QDEs inevitably
incorporate also collection efficiencies as the TB in front of the
detectors (and not beyond the crystal) is considered. The
reconstructed TB is characterized by the following values of its
parameters determined by Eqs.~(\ref{8}) and (\ref{6}): $ M_p = 38
$, $ b_p = 0.16 $, $ M_s = 1.4 \times 10^{-3} $, $ b_s = 39 $, $
M_i = 5.0 \times 10^{-3} $ and $ b_i = 24 $. Its joint
signal-idler PND $ p(n_s,n_i) $ is shown in Fig.~\ref{fig3}(b). On
average, the TB is composed of 6.2 photon pairs distributed over
38 modes, 0.05 signal noise photons and 0.11 idler noise photons.
This means that less than 2\% of photons in the beam are noise
photons. The numbers $ M_s $ and $ M_i $ of signal and idler noise
modes given by the fitting method are less than 1. This means that
the corresponding super-Gaussian PNDs have high probabilities at
the zero photon number and low but long tails. We attribute this
form of PNDs to distortions of weak electronic signals inside the
detection chain and not exactly set zeroes at the voltages axes.
This means that the originally assumed noise optical fields
considered as noise photo-electron signals contain an important
contribution from electronic noise that even changes the type of
their statistics. Covariance between the signal and idler photon
numbers is $ 0.76\pm 0.02 $. Noise reduction factor $ R $
quantifying sub-shot-noise correlations between the signal and
idler photon numbers \cite{PerinaJr2012a} equals $ 0.37\pm 0.02 $
indicating a larger role of noise photons in the TB.


The presented two-step analysis can be simplified provided that
the noise signal and idler fields are weak compared to that of
photon pairs and can be omitted. In this case, both steps can be
combined together giving the following formulas for the
theoretical [$ \langle n_p\rangle $, $ \langle (\Delta n_p)^2
\rangle $] and experimental [$ \langle v_s \rangle_a $, $ \langle
v_i \rangle_a $, $ \langle (\Delta v_s)^2 \rangle_a $, $ \langle
(\Delta v_i)^2 \rangle_a $, $ \langle \Delta v_s \Delta v_i
\rangle_a $] moments ($c=s,i$):
\begin{eqnarray}   
 \eta_c \delta v_c \langle n_p\rangle =
  \langle v_c\rangle_a , \hspace{2mm}
 \eta_s \delta v_s \eta_i \delta v_i \langle (\Delta n_p)^2 \rangle =
  \langle \Delta v_s \Delta v_i \rangle_a , \nonumber \\
 \eta_c^2 \delta v_c^2 \Biggl[ \langle (\Delta n_p)^2 \rangle
   + \frac{1-\eta_c}{\eta_c} \langle n_p\rangle \Biggr]
  = \langle (\Delta v_c)^2 \rangle_a . \hspace{10mm}
 \label{12}
\end{eqnarray}
Equations~(\ref{12}) represent five constrains for allowed values
of six unknown parameters. Two of them characterize the paired
field [$ \langle n_p\rangle $, $ \langle (\Delta n_p)^2 \rangle $]
whereas the remaining four parameters describe the detection
process. Relations in Eqs.~(\ref{12}) can be rewritten such that
the unknown parameters are expressed as functions of QDE $ \eta_s
$. An optimal value of QDE $ \eta_s $ is chosen, similarly as in
the general approach, such that the declination function $ D $ in
Eq.~(\ref{9}) minimizes.

Equations (\ref{12}) are rearranged as follows. First, the ratio
of equations for $ \langle v_s\rangle $ and $ \langle v_i\rangle $
provides the relation:
\begin{equation}  
 \eta_i \delta v_i = r \eta_s \delta v_s ,
\label{13}
\end{equation}
where $ r $ is defined in Eq.~(\ref{4}). By using $r$, the
equations for second moments in (\ref{12}) can be recast as:
\begin{eqnarray}   
 \eta_s^2 \delta v_s^2 \Biggl[ \langle (\Delta n_p)^2 \rangle
   + \frac{1-\eta_s}{\delta v_s \eta_s^2} \langle v_s\rangle_a \Biggr]
   &=& \langle (\Delta v_s)^2 \rangle_a , \nonumber \\
 \eta_s^2 \delta v_s^2 \Biggl[ \langle (\Delta n_p)^2 \rangle
   + \frac{1-\eta_i}{\delta v_s \eta_s \eta_i} \langle v_s\rangle_a \Biggr]
   &=& \frac{ \langle (\Delta v_i)^2 \rangle_a }{r^2} \nonumber \\
 &\equiv & \langle (\Delta v_i)^2 \rangle_r , \hspace{1cm} \nonumber \\
 \eta_s^2 \delta v_s^2 \langle (\Delta n_p)^2 \rangle &=&
  \frac{\langle \Delta v_s \Delta v_i \rangle_a}{r} \nonumber \\
 &\equiv & \langle \Delta v_s \Delta v_i \rangle_r .
 \label{14}
\end{eqnarray}
Combining the first and third equations in (\ref{14}) the
expression for $ \delta v_s $ as a function of QDE $ \eta_s $ can
be reached:
\begin{equation}   
 \delta v_s = \frac{ \langle (\Delta v_s)^2 \rangle_a - \langle
  \Delta v_s \Delta v_i \rangle_r}{ (1-\eta_s) \langle
  v_s\rangle_a} .
\label{15}
\end{equation}
Also the relation for QDE $\eta_i $ can be recovered:
\begin{equation}   
 \eta_i = \frac{ \left[ \langle (\Delta v_s)^2 \rangle_a - \langle
  \Delta v_s \Delta v_i \rangle_r \right] \eta_s }{ \langle (\Delta v_i)^2 \rangle_r - \langle
  \Delta v_s \Delta v_i \rangle_r +\eta_s \left[ \langle (\Delta v_s)^2 \rangle_a
  - \langle (\Delta v_i)^2 \rangle_r \right] } .
\label{16}
\end{equation}
The voltage window's width $ \delta v_i $ is then obtained as:
\begin{equation}   
 \delta v_i =  r \delta v_s \eta_s / \eta_i.
\label{17}
\end{equation}

The photon-number moments of the paired field can be easily
determined using Eqs.~(\ref{12}):
\begin{equation} 
 \langle n_p\rangle = \frac{ \langle v_s\rangle_a}{\eta_s \delta
  v_s} , \hspace{5mm}
 \langle (\Delta n_p)^2 \rangle = \frac{ \langle \Delta v_s \Delta v_i
  \rangle_a }{ \eta_s \delta v_s \eta_i \delta v_i} \; .
\label{18}
\end{equation}
Finally, the number $ M_p $ of modes and their average mean
photon-pair number $ b_p $ are obtained by using Eqs.~(\ref{6}).

The analysis of the data leading to Fig.~\ref{fig2} shows that the
declination function $ D $ decreases with QDE $ \eta_s $ in the
interval of allowed values $ \eta_s $. Also the remaining three
parameters, $ \eta_i $, $ \delta v_s $ and $ \delta v_i $, are
monotonous functions of the QDE $ \eta_s $ which makes the
optimization procedure stable.
The declination function $ D $ thus attains its minimum value at
the border, where $ \eta_s = 0.09\pm 0.005 $, $ \eta_i = 0.086 \pm
0.005 $, $ \delta v_s = 0.341 $, and $ \delta v_i = 0.422 $. The
TB contains on average 6.1 photon pairs found in $ M_p = 1800 $
independent modes ($ b_p = 3.3 \times 10^{-3} $). Comparison of
these results with the previous ones reveals that the simplified
method is able to determine QDEs with the relative precision
better than 5\% for the TB containing 2\% of noise photons. We
estimate that relative precision better than 10\% is reached
provided that the noise photons form less than 4\% of a TB. The
method also provides reasonable mean photon-pair numbers $ \langle
n_p \rangle $. On the other hand, it overestimates windows' widths
$ \delta v_s $ and $ \delta v_i $ and assigns larger numbers $ M_p
$ of independent photon-pair modes. This originates in the
presence of non-negligible amount of noise photons in the analyzed
TB. However, these drawbacks are acceptable and they are
outweighed by relative simplicity of the simplified method
compared to the general two-step approach. The simplified method
is also more suitable for the analysis of data collected with more
intense TBs.

In conclusion, we have developed and experimentally verified a
method allowing absolute detector calibration of
photon-number-resolving detectors with analog output using twin
beams. It gives detection efficiencies with high precision
provided that the noise is sufficiently small. Especially
electronic noise is not treated by the method in its full
complexity and its more rigorous description would give more
insight into measurement uncertainties and result in the improved
precision. The method also reveals parameters of twin beams. A
simplified approach suitable for more intense twin beams has been
suggested. Experimental conditions for its use have been found.

Support by projects P205/12/0382 of GA \v{C}R,
CZ.1.05/2.1.00/03.0058 and CZ.1.07/2.3.00/20.0058 of M\v{S}MT
\v{C}R and MIUR (FIRB LiCHIS - RBFR10YQ3H) are acknowledged.


\end{document}